\documentclass[12pt]{article}

\usepackage[centertags]{amsmath}
\usepackage{amsmath}
\usepackage{graphicx}

\newcommand{\RR}{{\mathbb R}}

\newcommand{\CC}{{\mathbb C}}
\newcommand{\beq}{\begin{equation}}
\newcommand{\eeq}{\end{equation}}
\newcommand{\ba}{\begin{array}}
\newcommand{\ea}{\end{array}}
\newcommand{\bea}{\begin{eqnarray}}
\newcommand{\eea}{\end{eqnarray}}

\usepackage[centertags]{amsmath}
\usepackage{amsfonts}
\usepackage{amssymb}
\usepackage{amsthm}
\usepackage{newlfont}
\usepackage{epsfig}
\usepackage{amscd}

\begin{document}

\begin{center}

{\large \sc \bf Integrable dispersionless PDEs arising as commutation condition of pairs of vector fields}

\vskip 20pt

{\large  S. V. Manakov$^{1,\S}$ and P. M. Santini$^{2,\S}$}

\vskip 20pt

{\it 
$^1$ Landau Institute for Theoretical Physics, Moscow, Russia

\smallskip

$^2$ Dipartimento di Fisica, Universit\`a di Roma "La Sapienza", and \\
Istituto Nazionale di Fisica Nucleare, Sezione di Roma 1 \\
Piazz.le Aldo Moro 2, I-00185 Roma, Italy}

\bigskip

$^{\S}$e-mail:  {\tt paolo.santini@roma1.infn.it}

\end{center}

\begin{abstract}

In this paper we review some results about the theory of integrable dispersionless PDEs arising as commutation condition of pairs of one-parameter families of vector fields, developed by the authors during the last years. We review, in particular, the basic formal aspects of a novel Inverse Spectral Transform including, as inverse problem, a nonlinear Riemann - Hilbert (NRH) problem, allowing one i) to solve the Cauchy problem for the target PDE; ii) to construct classes of RH spectral data for which the NRH problem is exactly solvable, corresponding to distinguished examples of exact implicit solutions of the target PDE; iii) to construct the longtime behavior of the solutions of such PDE; iv) to establish in a simple way if a localized initial datum breaks at finite time and, if so, to study analytically how the multidimensional wave breaks. We also comment on the existence of recursion operators and Backl\"und - Darboux transformations for integrable dispersionless PDEs.
\end{abstract}

\section{Introduction}
Waves propagating in weakly nonlinear and dispersive media are well described by integrable soliton equations, like the Korteweg - de Vries \cite{KdV}, the Nonlinear Scrh\"odinger \cite{ZSNLS} equations and their integrable $(2+1)$ dimensional generalizations, the Kadomtsev - Petviashvili \cite{KP} and Davey - Stewartson \cite{DS} equations respectively. The Inverse Spectral Transform (IST), introduced by Gardner, Green, Kruskal and Miura \cite{GGKM}, is the spectral method allowing one to solve the Cauchy problem for such PDEs, predicting that a localized disturbance evolves into a number of soliton pulses + radiation, and solitons arise as an exact balance between nonlinearity and dispersion \cite{ZMNP}-\cite{AC}. It is known that, apart from exceptional cases, soliton PDEs do not generalize naturally to more than $(2+1)$ dimensions; therefore, in the context of soliton equations, integrability is a property of low dimensional PDEs. There is another important class of integrable PDEs, the so-called dispersionless PDEs (dPDEs), or PDEs of hydrodynamic type, arising in various problems of Mathematical Physics and intensively studied in the recent literature (see, f.i., \cite{ZS} - \cite{KM2}). Since integrable dPDEs arise from the condition of commutation $[\hat L_1 ,\hat L_2 ]=0$ of pairs of one-parameter families of vector fields, implying the existence of common zero energy eigenfunctions:
\beq\label{Lax_pair}
[\hat L_1 ,\hat L_2 ]=0~~\Rightarrow~~\hat L_j\psi=0,~~j=1,2,
\eeq
they can be in an arbitrary number of dimensions \cite{ZS}. Due to the lack of dispersion, these multidimensional PDEs may or may not exhibit a gradient catastrophe at finite time, and their integrability gives a unique chance to study analytically such a mechanism. Also with this motivation, a novel IST for vector fields, significantly different from that of soliton PDEs \cite{ZMNP,AS,AC}, has been recently constructed \cite{MS0,MS1,MS2} i) to solve the Cauchy problem for dPDEs, ii) obtain the longtime behavior of solutions, iii) costruct distinguished classes of exact implicit solutions, iv) establish if, due to the lack of dispersion, the nonlinearity of the PDE is ``strong enough'' to cause the gradient catastrophe of localized multidimensional disturbances, and v) study analytically the breaking mechanism \cite{MS0}-\cite{MS8}. 

This paper is the written rendition of the talk presented by one of the authors (PMS) at the PMNP 2013 Conference, in a special session dedicated to the memory of Sergey V. Manakov, and it contains some of the results obtained by the authors in the period 2005-2011 (in \S 2 and \S 3), before the premature death of Sergey, with some additional formulas and considerations (in \S 4) not written before. This paper is dedicated to the memory of Sergey, a great scientist and a loyal friend.    

\section{Two basic examples}
Here we consider, as illustrative examples, two integrable dPDEs; the first is associated with a Lax pair of vector fields not containing the partial derivative with respect to the spectral parameter $\lambda$, the second containing it.\\
\\
1. The N-vector PDE in $(N+4)$ dimensions \cite{MS1}:
\beq\label{MS_SDYM}
\ba{l}
\vec u_{t_1z_2}-\vec u_{t_2z_1}+
\left(\vec u_{z_1}\cdot\nabla_{\vec x}\right)\vec u_{z_2}-\left(\vec u_{z_2}\cdot\nabla_{\vec x}\right)\vec u_{z_1}=\vec 0,\\
\ea
\eeq
equivalent to the commutativity condition $[\hat L_1 ,\hat L_2 ]=\hat 0$ of the $(N+1)$ dimensional vector fields
\beq\label{Lax_MS_SDYM}
\ba{l}
\hat L_i=\partial_{t_i}+\lambda\partial_{z_i}+\vec u_{z_i}\cdot\nabla_{\vec x},~~~~i=1,2, 
\ea
\eeq
where $\vec u(t_1,t_2,z_1,z_2,\vec x)\in \RR^N$, $\vec x=(x_1,\dots,x_N)\in \RR^N$ and $\nabla_{\vec x}=(\partial_{x_1},..,\partial_{x_N})$, together with its deepest Hamiltonian reduction, the scalar PDE in $(2M+4)$ dimensions (for $N$ even and $M=N/2$) and its Hamiltonian Lax pair \cite{San1}:
\beq\label{scalar_Ham_multidim}
\ba{l}
\theta_{t_2 z_1}-\theta_{t_1 z_2}+\{\theta_{z_1},\theta_{z_2}\}_{\vec x}=c(t_1,t_2,z_1,z_2), \\
\hat L_j =\partial_{t_j}+\lambda\partial_{z_j}+\{\theta_{z_j},\cdot\},~~j=1,2 , \\
\{f,g\}_{\vec x}\equiv \sum\limits_{k=1}^M(f_{x_k}g_{x_{M+k}}-f_{x_{M+k}}g_{x_k}),
\ea
\eeq
where $c$ is an arbitrary function of its arguments, reducing, for $N=2$, to the well-known first and second heavenly equations of Plebanski \cite{Plebanski}  
\beq\label{heavenly}
\ba{l}
\{\theta_{z_1},\theta_{z_2}\}_{x_1,x_2}=c(z_1,z_2),~~~\mbox{first heavenly equation} \\
\theta_{t_2 x_1}-\theta_{t_1 x_2}+\{\theta_{x_1},\theta_{x_2}\}_{x_1,x_2}=c(t_1,t_2),~~~\mbox{second heavenly equation},
\ea
\eeq
obtained assuming respectively that $\theta$ does not depend on $t_j,j=1,2$ and that $z_j=x_j,j=1,2$. \\
\\
2. The following system of two nonlinear PDEs in $2+1$ dimensions and its Lax pair of two - dimensional vector fields \cite{MS2}:
\beq
\label{dKP-system}
\ba{l}
u_{xt}+u_{yy}+(uu_x)_x+v_xu_{xy}-v_yu_{xx}=0,         \\
v_{xt}+v_{yy}+uv_{xx}+v_xv_{xy}-v_yv_{xx}=0, \\
\hat L_1\equiv \partial_y+(\lambda+v_x)\partial_x-u_x\partial_{\lambda}, \\
\hat L_2\equiv \partial_t+(\lambda^2+\lambda v_x+u-v_y)\partial_x+(-\lambda u_x+u_y)\partial_{\lambda}, 
\ea
\eeq
describing a fairly general Einstein - Weyl metric \cite{Dunajj}. This system reduces, for $v=0$, to the celebrated dispersionless Kadomtsev - Petviashvili (dKP) equation and its Lax pair of Hamiltonian two-dimensional vector fields:
\beq\label{dKP}
\ba{l}
u_{xt}+u_{yy}+(uu_x)_x=0,~~u=u(x,y,t)\in\RR,~~x,y,t\in\RR, \\
\hat L_1\equiv \partial_y+\lambda\partial_x-u_x\partial_{\lambda}=
\partial_y+\lambda\partial_x+\{\frac{\lambda^2}{2}+u,\cdot \}_{\lambda,x}, \\
\hat L_2\equiv \partial_t+(\lambda^2+u)\partial_x+(-\lambda u_x+u_y)\partial_{\lambda}=\partial_t+\{\frac{\lambda^3}{3}+\lambda u-\partial^{-1}_xu_y,\cdot\}_{\lambda,x}, \\
\{ f, g \}_{\lambda,x}\equiv f_{\lambda}g_t - f_t g_{\lambda} ,
\ea
\eeq
describing the evolution of weakly nonlinear, nearly one-dimensional waves in Nature, in the absence of dispersion and dissipation \cite{Timman},\cite{ZK},\cite{MS8}, and, for $u=0$, to the Pavlov equation \cite{Pavlov} and its Lax pair \cite{Duna} of non Hamiltonian one-dimensional vector fields:
\beq 
\label{Pavlov}
\ba{l}
v_{xt}+v_{yy}+v_xv_{xy}-v_yv_{xx}=0,~~v=v(x,y,t)\in\RR,~~x,y,t\in\RR, \\
\hat L_1\equiv \partial_y+(\lambda +v_x)\partial_x, \\
\hat L_2\equiv \partial_t+(\lambda^2+\lambda v_x-v_y)\partial_x.
\ea
\eeq
\section{The IST for vector fields}

Since the Lax pair of integrable dPDEs is made of $n$-dimensional vector fields, Hamiltonian in some cases, the zero - energy eigenfunctions satisfy the following basic properties.\\
1) {\it The space of eigenfunctions is a ring}.  If $f_1,f_2,\dots$ are solutions of the Lax pair equations (\ref{Lax_pair}), then an arbitrary differentiable function $F(f_1,,f_2,\dots)$ of them is also a solution of (\ref{Lax_pair}). A basis of this ring consists of $n$ independent eigenfunctions. \\   
2) {\it In the Hamiltonian reduction, the space of eigenfunctions is also a Lie algebra, whose Lie bracket is the natural Poisson bracket}. If $f_1,f_2,\dots$ are solutions of the Lax pair (\ref{Lax_pair}), then the Poisson bracket of any two of them $\{f_i,f_j\}$ is also a solution of (\ref{Lax_pair}). \\
For the sake of concreteness, we concentrate on the IST for the system (\ref{dKP-system}) \cite{MS0,MS1,MS2,MS3}. Assuming that $\vec u=(u,v)^T\to \vec 0$ as $x^2+y^2\to\infty$, we have that $\vec\xi=(\xi,\lambda)^T,~\xi\equiv x-\lambda y$ is the basic eigenfunction of $\hat L_1$ at $x^2+y^2\to\infty$. We also assume that $\vec u=(u,v)^T\in\RR^2$; it follows that, for $\lambda\in\RR$, the vector fields are real. \\
\\
\textbf{The Direct Problem} A basic role in the IST for real vector fields is played by a suitable basis of real (Jost) eigenfunctions $\vec\phi_{\pm}(x,y,\lambda)$, defined for $\lambda\in\RR$, the solutions of equations $\hat L_1 \vec\phi_{\pm}=\vec 0$ satisfying the boundary conditions 
\beq
\vec\phi_{\pm}(x,y,\lambda) \to  \vec\xi,~~y\to\pm\infty,~~\lambda\in\RR, 
\eeq 
intimately related to the system of real ODEs 
\beq\label{ODE}
\frac{dx}{dy}=\lambda+v_x(x,y),~~\frac{d\lambda}{dy}=-u_x(x,y)
\eeq
defining the characteristics of $\hat L_1$. If the potentials $(u,v)$ are sufficiently regular, the solution $(x(y),\lambda(y))$ of the ODE (\ref{ODE}) exists unique globally in the (time) variable $y$, with the following free particle asymptotic behavior
\beq\label{asympt_ODE_dKP}
x(y)\to \lambda_{\pm}y+x_{\pm},~~\lambda(y)\to \lambda_{\pm},~~y\to\pm\infty,
\eeq 
reducing to the asymptotics
\beq\label{asympt_ODE_Pavlov}
x(y)\to \lambda y+x_{\pm},~~\lambda(y)=\lambda=\mbox{constant},~~y\to\pm\infty,
\eeq 
in the Pavlov reduction $u=0$. Once the asymptotics $\lambda_{\pm},x_{\pm}$ are constructed in terms of the initial data $x_0=x(y_0),\lambda_0=\lambda(y_0)$ of the ODE: $\lambda_{\pm}(x_0,y_0,\lambda_0),x_{\pm}(x_0,y_0,\lambda_0)$, the real eigenfunctions $\vec\phi_{\pm}$, that are particular constants of motion of the ODE, are given by
\beq\label{asympt_ODE_Pavlov}
\vec\phi_{\pm}(x_0,y_0,\lambda_0)=(x_{\pm}(x_0,y_0,\lambda_0),\lambda_{\pm}(x_0,y_0,\lambda_0)).
\eeq 
Another important ingredient of the formalism is given by the complex eigenfunction $\vec\psi$, defined by the asymptotics
\beq
\vec\psi(y,\vec x,\lambda)\sim \vec\xi,~~x^2+y^2\to\infty ,~~\lambda\notin\RR,
\eeq
analytic for $\lambda\notin\RR$, having continuous boundary values $\vec\psi^{\pm}(x,y,\lambda),~\lambda\in\RR$ from above and below the real $\lambda$ axis, with the following asymptotics for large complex $\lambda$:
\beq
\label{asympt-psi}
\ba{l}
\vec\psi^{\pm}(x,y,\lambda)=\vec \xi+\frac{1}{\lambda}\vec U(x,y)+\vec O\left(\frac{1}{\lambda^2}\right),~~|\lambda |>>1, \\
\vec U(x,y)\equiv \left(
\ba{c}
-yu(x,y)-v(x,y) \\
u(x,y)
\ea
\right).
\ea
\eeq
The analyticity properties of $\vec\psi^{\pm}$ and of their $y\to\pm\infty$ limits follow from those of the analytic Green's functions $G^{\pm}$ of the undressed operator $\partial_y+\lambda\partial_x$
\beq\label{Green_analytic}
G^{\pm}(x,y,\lambda)=\pm\left(2\pi i(x-(\lambda \pm i\epsilon) y)\right)^{-1},
\eeq
exhibiting the following asymptotics for $y\to\pm\infty$:
\beq\label{G_asymt}
\ba{l}
G^{\pm}(x-x',y-y',\lambda)\to \pm \left(2\pi i(\xi-\xi'\mp i\epsilon)\right)^{-1},\;\;y\to +\infty, \\
G^{\pm}(x-x',y-y',\lambda)\to \pm \left(2\pi i(\xi-\xi'\pm i\epsilon)\right)^{-1},\;\;y\to -\infty,
\ea
\eeq
and entailing that: the $y=+\infty$ asymptotics of $\vec\psi^{+}$ and $\vec\psi^{-}$ are analytic respectively in the lower and upper halves of the complex plane $\xi$, while the $y=-\infty$ asymptotics of $\vec\psi^{+}$ and $\vec\psi^{-}$ are analytic respectively in the upper and lower halves of the complex plane $\xi$. Similar features have been observed first in \cite{MZ}. In addition $\vec\psi^{\pm}-\vec\xi=\vec O(\xi^{-1})$ as $|\xi |\gg 1$. \\
\\
\textbf{Scattering and spectral data}. The $y=+\infty$ limit of $\vec\phi_-$ defines the natural ($y$ - time) {\it scattering vector} $\vec\sigma$ for $\hat L_1$:
\beq
\label{def-S}
\displaystyle\lim_{y\to +\infty}\vec\phi_-(x,y,\lambda) \equiv 
\vec{\cal S}(\vec \xi)=\vec \xi+\vec\sigma(\vec \xi).
\eeq
Since the space of eigenfunctions is a ring, the eigenfunctions $\vec\psi^{\pm}$ for $\lambda\in\RR$ can be expressed in terms of the real eigenfunctions $\vec\phi_{\pm}$, and this expression defines the {\it spectral data} $\vec\chi^{\pm}_{\beta}(\xi,\lambda)$:
\beq\label{psi_phi}
\ba{l}
\vec\psi^{\pm}(x,y,\lambda)=\vec{\cal K}^{\pm}_{-}(\vec\phi_-(x,y,\lambda))=\vec{\cal K}^{\mp}_{+}(\vec\phi_+(x,y,\lambda)),~~\lambda\in\RR,  \\
\vec{\cal K}^{\pm}_{\beta}(\vec\xi)\equiv \vec\xi+\vec\chi^{\pm}_{\beta}(\vec\xi),~~\vec\xi =(\xi,\lambda),
\ea
\eeq
where $\vec\chi^{+}_{\beta}(\vec\xi)$ and $\vec\chi^{-}_{\beta}(\vec\xi)$ are analytic wrt the first argument $\xi$ respectively in the upper and lower halves of the complex $\xi$ - plane, as a consequence of equations (\ref{G_asymt}) and of the above analyticity properties of $\vec\psi^{\pm}$ as $y\to\pm\infty$. We remark tha equations (\ref{psi_phi}) for $\vec\psi^{-}$ can be omitted taking account of the symmetry properties coming from the reality of the potentials:
\beq\label{reality}
(u,v)\in\RR^2~\Rightarrow~ \vec\phi_{\pm}\in\RR^2,~\vec\psi^-=\overline{\vec\psi^+},~~\vec\sigma\in\RR^2,~\vec\chi^-_{\alpha}=\overline{\vec\chi^+_{\alpha}},~~\lambda\in\RR . 
\eeq
Evaluating the second of equations (\ref{psi_phi}) for $\vec\psi^{+}$ at $y=+\infty$, one obtains the following linear Riemann - Hilbert (RH) problem with a shift:
\beq\label{RH_shift}
\ba{l}
\vec\sigma(\xi,\lambda)+\vec\chi^+_-(\vec\xi +\vec\sigma(\xi,\lambda))-\vec\chi^-_+(\xi ,\lambda)=\vec 0, \\
|\vec\chi^{\pm}_{\beta}(\xi,\lambda)|=O(\xi^{-1}),~~\xi\sim\infty
\ea
\eeq
equivalent to a linear Fredholm integral equation \cite{Gakhov}, allowing one to uniquely construct the spectral data $\vec\chi^{+}_-$ and $\vec\chi^{-}_+$ from the scattering data $\vec\sigma$, under the hypothesis that the mapping $\xi\to \xi+\sigma_1(\xi,\lambda)$ be invertible. \\
Soon after the introduction of the RH problem (\ref{RH_shift}) in \cite{MS0}, an alternative construction was proposed, based on the linear integral equations \cite{MS1,MS2}
\beq
\label{Fourier-varphi-psi}
\ba{l}
\tilde{\vec\chi}^+_-(\vec\omega)+\theta(\omega_1)\left(\tilde{\vec\sigma}(\vec\omega)+
\int_{\RR^2}\tilde{\vec\chi}^+_-(\vec\eta)Q(\vec\eta,\vec\omega)d\vec\eta\right)=\vec 0,  \\
\tilde{\vec\chi}^-_-(\vec\omega)+\theta(-\omega_1)\left(\tilde{\vec\sigma}(\vec\omega)+
\int_{\RR^2}\tilde{\vec\chi}^-_-(\vec\eta)Q(\vec\eta,\vec\omega)d\vec\eta\right)=\vec 0,
\ea
\eeq 
involving the Fourier transforms  $\tilde{\vec\sigma}$ and $\tilde{\vec\chi}^{\pm}_{\beta}$ of $\vec\sigma$ and ${\vec\chi}^{\pm}_{\beta}$:
\beq
\label{Fourier-sigma}
\tilde{\vec\sigma}(\vec\omega)=\int_{\RR^2}\vec\sigma(\vec\xi)e^{-i\vec\omega\cdot\vec\xi}d\vec\xi,~~~
\tilde{\vec\chi}^{\pm}_{\beta}(\vec\omega)=\int_{\RR^2}{\vec\chi}^{\pm}_{\beta}(\vec\xi)e^{-i\vec\omega\cdot\vec\xi}d\vec\xi
\eeq
and the scalar kernel:
\beq
\label{def-Q}
Q(\vec\eta,\vec\omega)=\frac{1}{(2\pi)^2}\int_{\RR^2}e^{i(\vec\eta-\vec\omega)\cdot\vec\xi}[e^{i\vec\eta\cdot\vec\sigma(\vec\xi)}-1]d\vec\xi.
\eeq 
To construct, say, (\ref{Fourier-varphi-psi}) for $\tilde{\vec\chi}^+_- $, one applies the integral operator $\int_{\RR}d\xi e^{-i\vec\omega\cdot\vec\xi}\cdot$ for $\omega_1>0$ to equation (\ref{RH_shift}), using the above analyticity properties and the Fourier representations of ${\vec\chi}^{\pm}$ and $\vec\sigma$. 

We remark that, in the Pavlov reduction $u=0$, ${\phi_2}_{\pm}={\psi^{\pm}_2}=\lambda$, implying that $\sigma_2=\chi^{\pm}_2=0$, while in the Hamiltonian dKP reduction $v=0$ \cite{MS2}: 
\beq
\{{\phi_{\pm}}_1,{\phi_{\pm}}_2 \}_{x,\lambda}=\{\psi^{\pm}_1,\psi^{\pm}_2 \}_{x,\lambda}=1,
\eeq
and, consequently:
\beq
\ba{l}
\{{\cal S}_1,{\cal S}_2 \}_{\vec\xi}=\{{\cal K}^{\pm}_1,{\cal K}^{\pm}_2 \}_{\vec\xi}=1.
\ea
\eeq
 
Summarizing, from the initial data $\vec u(x,y,0)=(u(x,y,0),v(x,y,0))$ one constructs the real eigenfunctions $\vec\phi_-$ and then the scattering data $\vec\sigma(\xi,\lambda)$ through the solution of the ODE system (\ref{ODE}). From the scattering data one constructs the spectral data ${\vec\chi}^{\pm}$ through the solution of the RH problem with shift (\ref{RH_shift}). The main difficulties to make the above formalism rigorous are associated with the proof of the existence of the analytic eigenfunctions, and of their limits on the real $\lambda$ axis from above and below. While such a proof exists \cite{GSW} for the Pavlov reduction $u=0$, for the dKP reduction the existence of the analytic eigenfunctions is proven, at the moment, only sufficiently far from the real $\lambda$ axis \cite{GS}. We remark that, while the construction of the spectral data from the scattering data through the RH problem with shift (\ref{RH_shift}) does not present difficulties, the one that makes use of the integral equations (\ref{Fourier-varphi-psi}) may not be easy, due to the behavior of its kernel, as pointed out in \cite{GSW}. \\ 
\\
\textbf{Two inverse problems} The first inversion (the reconstruction of $\vec\phi_-$ from the spectral data $\vec\chi^+_-$) is provided by the nonlinear integral equation 
\beq\label{inversion1}
\vec\phi_-(x,y,\lambda)+H_{\lambda}\vec\chi^+_{-I}(\vec\phi_-(x,y,\lambda)+\vec\chi^+_{-R}(\vec\phi_-(x,y,\lambda))=\vec\xi,
\eeq
where $\vec\chi^+_{-R}$ and $\vec\chi^+_{-I}$ are the real and imaginary parts of $\vec\chi^+_-$ , and $H_{\lambda}$ is the Hilbert transform operator wrt $\lambda$
\beq
H_{\lambda}f(\lambda)=\frac{1}{\pi}PV \int\limits_{-\infty}^{\infty}\frac{f(\lambda')}{\lambda-\lambda'}d\lambda' .
\eeq
Since $\vec\chi^+_-(\xi,\lambda)$ is analytic wrt $\xi$, its real and imaginary parts must satisfy the relation $\vec\chi^+_{-R}+H_{\xi}\vec\chi^+_{-I}=\vec 0$. Equation (\ref{inversion1}) expresses the fact that the RHS of (\ref{psi_phi}) for $\vec\psi^{+}$ is the boundary value of a function analytic in the upper half $\lambda$ plane.

Once $\vec\phi_-$ is reconstructed from $\vec\chi^{+}_-$ solving the nonlinear integral equation (\ref{inversion1}), equations (\ref{psi_phi}) give $\vec\psi^{\pm}$, and $(u,v)$ is finally reconstructed from 
\beq\label{inv_u_1}
\ba{l}
u(x,y)=\displaystyle\lim_{\lambda\to\infty}{\left(\lambda(\psi^-_2(x,y,\lambda)-\lambda \right)}, \\
v(x,y)=-yu-\displaystyle\lim_{\lambda\to\infty}{\left(\lambda(\psi^-_1(x,y,\lambda)-(x-\lambda y)\right)}, 
\ea
\eeq
consequence of (\ref{asympt-psi}). This inversion procedure was first introduced in \cite{Manakov1}. 

A second inverse problem can be obtained eliminating the real eigenfunctions from the first of equations (\ref{psi_phi}) for $\vec\psi^{\pm}$, obtaing a 2 vector nonlinear RH (NRH) problem on the real line:
\beq\label{NRH1}
\ba{l}
\psi^+_1(\lambda)={\cal R}_1\left(\psi^-_1(\lambda),\psi^-_2(\lambda)\right),~~\lambda\in\RR, \\
\psi^+_2(\lambda)={\cal R}_2\left(\psi^-_1(\lambda),\psi^-_2(\lambda)\right), \\
\psi^+_1(\lambda)=-y\lambda +x+O(\lambda^{-1}),~~\psi^+_2(\lambda)=\lambda +O(\lambda^{-1}),~~\lambda\sim\infty .
\ea
\eeq
or, in vector form: 
\beq\label{NRH2}
\vec\psi^+(\lambda)=\vec{\cal R}\left(\vec\psi^- (\lambda)\right),~~\lambda\in\RR,
\eeq
for the RH data $\vec{\cal R}$, constructed, via algebraic manipulation, from the spectral data. Once the analytic eigenfunctions are reconstructed through the solution of the NRH problem (\ref{NRH1}), the solution of the nonlinear PDE (\ref{dKP-system}) is obtained from (\ref{inv_u_1}). 

We remark that, in the two basic reductions, the RH data are constrained as follows:
\beq\label{constraints}
\ba{l}
{\cal R}_2(\zeta_1,\zeta_2)=\zeta_2,~~~\mbox{Pavlov reduction,} \\
\{{\cal R}_1,{\cal R}_2\}_{\zeta_1,\zeta_2}=1,~~~\mbox{dKP reduction.}
\ea
\eeq
\textbf{Evolution of the spectral data}. The evolution of the scattering, spectral, and RH data is described by the following simple formula \cite{MS2,MS3}:
\beq
\ba{l}
\Sigma_1(\xi,\lambda,t)=\Sigma_1(\xi-\lambda^2 t,\lambda,0)
\ea
\eeq
for the Pavlov equation, and 
\beq
\ba{l}
\Sigma_1(\xi,\lambda,t)=t\left(\Sigma_2(\xi-\lambda^2t,\lambda,0)\right)^2+\Sigma_1(\xi-\lambda^2t,\lambda,0), \\
\Sigma_2(\xi,\lambda,t)=\Sigma_2(\xi-\lambda^2t,\lambda,0)
\ea
\eeq
for the dKP equation. 

We remark that, from the eigenfunctions $\vec\phi_{\pm},\vec\psi^{\pm}$ of $\hat L_1$, one can constructs the common eigenfunctions $\vec\Phi_{\pm},\vec\Psi^{\pm}$ of $\hat L_1$ and $\hat L_2$ through the formulae
\beq
\ba{l}
{\Phi_{\pm}}_1={\phi_{\pm}}_1-t\left({\phi_{\pm}}_2\right)^2,~~{\Phi_{\pm}}_2={\phi_{\pm}}_2, \\
\Psi^{\pm}_1=\psi^{\pm}_1-t\left(\psi^{\pm}_2\right)^2,~~\Psi^{\pm}_2=\psi^{\pm}_2,
\ea
\eeq
and the inverse (dressing) problem for the common eigenfunctions reads as follows. \\
\\
\textbf{Nonlinear Riemann - Hilbert dressing} \cite{MS2,MS3,MS4}. Let $\vec\Psi^{\pm}(\lambda)$ be the solutions of the following 2 vector NRH problem on the line
\beq\label{NRH3}
\ba{l}
\vec\Psi^+(\lambda)=\vec{\cal R}\left(\vec\Psi^- (\lambda)\right),~~\lambda\in\RR, 
\ea
\eeq
with the normalization
\beq\label{normalization}
\ba{l}
\vec\Psi^{\pm}(\lambda)=
\left(
\ba{c}
-t\lambda^2-y\lambda +x-2ut \\
\lambda
\ea
\right)
+\vec O (\lambda^{-1}),~~\lambda\sim\infty ,
\ea
\eeq
for the RH data $\vec{\cal R}(\vec\zeta)=({\cal R}_1(\vec\zeta),{\cal R}_2(\vec\zeta)),~ \vec\zeta\in\CC^2$. Then $\vec\Psi^{\pm}(\lambda)$ are eigenfunctions of $\hat L_j~j=1,2$: $\hat L_j\vec\Psi^{\pm}=\vec 0,~j=1,2$, and $(u,v)$, constructed through the analogue of formulae (\ref{inv_u_1}):
\beq\label{inv_u_2}
\ba{l}
u(x,y)=\displaystyle\lim_{\lambda\to\infty}{\left(\lambda\left[\Psi^-_2(x,y,\lambda)-\lambda \right]\right)}, \\
v(x,y)=-yu-\lim\limits_{\lambda\to\infty}{\left(\lambda\left[\Psi^-_1+t(\Psi^{\pm}_2)^2-(x-\lambda y)\right]\right)}, 
\ea
\eeq
are solutions of the nonlinear system (\ref{dKP-system}). 

If the spectral data $\vec {\cal R}(\vec \zeta)$ satisfy the reality constraint
\beq\label{R_reality}
\ba{l}
\vec{\cal R}(\overline{ \vec{\cal R}(\bar{\vec\zeta})})=\vec\zeta,
~~~\forall\vec\zeta\in\CC^2,
\ea
\eeq
then the solutions $u,v$ are real: $u,v\in \RR$. \\
If the transformation $\vec \zeta \to \vec {\cal R}(\vec \zeta)$ is simplectic:
\beq\label{R_dKP}
\ba{l}
\{{\cal R}_1,{\cal R}_2\}_{\zeta_1,\zeta_2}=1,
\ea
\eeq
then $v=0$, $\vec\Psi^{\pm}(\lambda)$  are common eigenfunctions of the Hamiltonian vector fields (\ref{dKP}), and $u$ satisfies the dKP equation (\ref{dKP}). \\
If ${\cal R}_2(\zeta_1,\zeta_2)=\zeta_2$, then $\Psi^{\pm}_2(\lambda)=\lambda$, $u=0$ and $v$ satisfies the Pavlov equation (\ref{Pavlov}).

From the integral equations 
\beq\label{pi_equ}
\ba{l}
\Psi^-_1(\lambda)=-\lambda^2t-\lambda y+x-2ut+
\frac{1}{2\pi i}\int\limits_{\RR}\frac{d\lambda'}{\lambda'-(\lambda -i0)}
R_1\Big(\Psi^-_1(\lambda'),\Psi^-_2(\lambda')\Big), \\
\Psi^-_2(\lambda)=\lambda+\frac{1}{2\pi i}\int\limits_{\RR}\frac{d\lambda'}{\lambda'-(\lambda -i0)}
R_2\Big(\Psi^-_1(\lambda'),\Psi^-_2(\lambda')\Big), \\
{\cal R}_j(\zeta_1,\zeta_2)\equiv \zeta_j+R_j(\zeta_1,\zeta_2),~~~j=1,2
\ea
\eeq 
characterizing the solutions of the RH problem (\ref{NRH3}), and from the definition (\ref{inv_u_2}),   
one obtains the following spectral characterization of the solution $u$:
\beq\label{inverse_1}
u=F\left(x-2ut,y,t\right)\in\RR,
\eeq
where the spectral function $F$, defined by 
\beq\label{inverse_2}
\ba{l}
F\left(\xi,y,t \right)=-\int\limits_{\RR}\frac{d\lambda}{2\pi i}R_2
\Big(\Psi^-_1(\lambda;\xi,y,t),\Psi^-_2(\lambda;\xi,y,t)\Big),
\ea
\eeq
is connected to the initial data via the direct problem \cite{MS2}. 

We remark that, while the IST of most of soliton PDEs provides a spectral representation of the solution involving, as parameters, the space-time coordinates, the inverse problem of system (\ref{dKP-system}) provides a spectral representation (\ref{inverse_1}), (\ref{inverse_2}) of the solution involving, as parameter, also the solution $u$ itself, in the combination $(x-2ut)$. This is the spectral mechanism for the breaking of a generic localized initial condition at finite time $t$. This spectral mechanism is absent for the Pavlov equation (\ref{Pavlov}) ($u=0$), having the same linear part as dKP, but different nonlinear part. 

The NRH dressing provides a very convenient tool i) to study the longtime behavior of the solutions of an integrable dispersionless PDE \cite{MS4,MS5,MS6}; ii) to easily detect if a localized initial disturbance evolving according to such a PDE goes through a gradient catastrophe at finite time (f.i., there is no gradient catastrophe for the second heavenly equation (\ref{heavenly}) \cite{MS1,MS5}, while there is a gradient catastrophe \cite{MS6} for the dispersionless 2D Toda (or Boyer - Finley) equation $u_{xx}+u_{yy}=(exp(u))_{tt}$ \cite{FP,BF}); iii) to investigate analytically the wave breaking mechanism of such multidimensional waves (see Figure 1) \cite{MS4,MS6}; iv) to construct classes of RH data giving rise to exactly solvable NRH problems, and to distinguished exact implicit solutions of the dispersionless PDEs through an algorithmic approach \cite{MS7,BDM,MS4,MS5,MS6}; v) to detect integrable differential reductions of the associated hierarchy of PDEs \cite{Bogdanov1,Bogdanov2}, like the Dunajski interpolating equation $v_{xt}+v_{yy}+cv_xv_{xx}+v_xv_{xy}-v_yv_{xx}=0$ \cite{Dunajj}, an integrable PDE interpolating between the dKP and the Pavlov equations, corresponding to the reduction $u=c v_x$ of system (\ref{dKP-system}).   

We also remark that the mechanism responsible for this feature, in our example, is that the vector field $\hat L_2$ is quadratic in $\lambda$ and, at the same time, it contains the partial differential operator $\partial_{\lambda}$. Due to that, the unknown field $u$ is present in the normalization of the eigenfunctions of the vector field. It is easy to see that these properties are shared by the whole dKP hierarchy, associated with time operators involving higher powers of $\lambda$.
\begin{figure}[h]
\includegraphics[width=14pc]{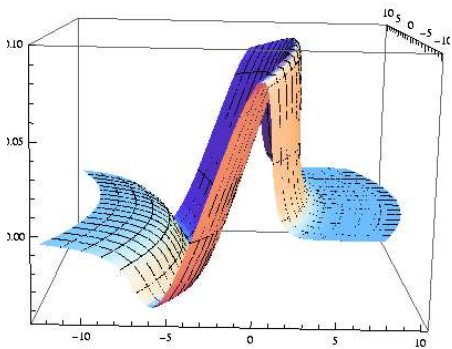}\hspace{2pc}%
\begin{minipage}[b]{14pc}\caption{\label{label} A small localized profile evolves according to the dKP equation into a parabolic wave front and breaks in a point of it. A detail of the parabolic wave front generated by a Gaussian initial profile around the breaking point at the breaking time, obtained from the analytical formula describing the wave breaking mechanism in the longtime regime \cite{MS4,MS8}.}
\end{minipage}
\end{figure}

\section{Recursion operators and hierarchies of dispersionless PDEs}
Soliton PDEs like the KdV and NLS equations possess infinitely many symmetries (and constants of motions), organized in integrable hierarchies of PDEs. These commuting flows are generated by Nijenhuis (or hereditary) operators, and possess a multi-Hamiltonian structure \cite{Magri, FF}. This pictures generalizes to integrable soliton PDEs in $(2+1)$ dimensions \cite{SF}. Soliton equations possess also discrete symmetries: the Backl\"und - Darboux transformations, allowing one to construct solutions from solutions through a recursive procedure \cite{AS, CD}. Do we have a similar picture for multidimensional dispersionless PDEs?

The picture is the exactly same for example (\ref{MS_SDYM}). We first observe that the vector equation (\ref{MS_SDYM}) and its Lax operators (\ref{Lax_MS_SDYM}) are nothing but the self-dual Yang - Mills (SDYM) equation \cite{AC} and its integrability structure
\beq\label{SDYM}
\ba{l}
~~[L_1,L_2]=0 ~~\Rightarrow~~U_{t_1 z_2}-U_{t_2 z_1}+[U_{z_1},U_{z_2}]=0, \\
L_i\equiv \partial_{t_i}+\lambda\partial_{z_i}+U_{z_i},~~i=1,2,
\ea
\eeq
corresponding to the Lie algebra of $N$ dimensional and $\lambda$ independent vector fields:
\beq\label{realization}
U \to \vec u \cdot\nabla_{\vec x},~~\vec u=\vec u(\vec x,t_1,t_2,z_1,z_2).
\eeq
Since the SDYM equation plays a role of master equation in the theory of integrable soliton PDEs \cite{AC}, equation (\ref{MS_SDYM}) should play an analogous role in the theory of integrable multidimensional dPDEs arising as commutativity condition of pairs of one parameter families of vector fields not containing derivatives with respect to the parameter $\lambda$ \cite{San1}.

It is known that the recursion operator \cite{BLR}:
\beq\label{RecOp_SDYM}
\ba{l}
R=\Theta_2 \left(\Theta_1\right)^{-1},~~~(\mbox{or}~~R'=\left(\Theta_1\right)^{-1}\Theta_2),
\ea
\eeq
defined as factorization of the bi-Hamiltonian structure
\beq\label{HamSDYM}
\ba{l}
\Theta_1=\partial_{z_1},~~\Theta_2=\partial_{t_1}+ad(U_{z_1}), \\
ad(g)f\equiv [g,f],
\ea
\eeq
generates the hierarchy of symmetries of the SDYM equation. Since this operator contains the $ad$ operator, it is perfectly well defined also in the vector field case (\ref{realization}) (the commutator of two vector fields is a vector field), giving rise to the following recursion operator and bi-Hamiltonian structures of the PDE (\ref{MS_SDYM}):
\beq\label{RecOp_MS-SDYM}
\ba{l}
\check R'=\left(\check \Theta_1\right)^{-1}\check \Theta_2, \\
\check \Theta_1=\partial_{z_1},~~\check \Theta_2 \vec f=
\vec f_{t_1}+\left(\vec u_{z_1}\cdot\nabla_{\vec x}\right)\vec f-\left(\vec f\cdot\nabla_{\vec x}\right)\vec u_{z_1}
\ea
\eeq
Also in this vector field setting, $\check \Theta_1,~\check \Theta_2$ is a compatible pair of Hamiltonian operators and $\check R$ is a Nijenhuis or hereditary operator, generating commuting symmetries of the PDE (\ref{MS_SDYM}). The recursion operator $\check R'$ in (\ref{RecOp_MS-SDYM}) was first derived in \cite{Marv_Sergyeyev} through different considerations (see also \cite{San1} for a third derivation of (\ref{RecOp_MS-SDYM}) and of the hierarchy of integrable flows associated with (\ref{MS_SDYM})).

If we try, instead, to construct the Backl\"und transformations of (\ref{MS_SDYM}) starting from the recursion operator $\tilde R$ of the Backl\"und transformations of the SDYM equation \cite{BLR}:
\beq\label{RecOp_Backlund_SDYM}
\ba{l}
\tilde R=\tilde \Theta_2 \left(\Theta_1\right)^{-1}, \\
\Theta_1=\partial_{z_1},~~\tilde \Theta_2 =\partial_{t_1}+\tilde{ad} (U_{z_1}) 
\ea
\eeq
depending on the generalized commutator
\beq
\tilde{ad} (U_{z_1})f\equiv \tilde U_{z_1} f-f U_{z_1} 
\eeq
(where $U$ and $\tilde U$ are two solutions of SDYM), one fails, since this generalized commutator of vector fields is not a vector field. For the same reason, one cannot use the construction of the Darboux transformations of the SDYM equation, defined by the generalized commutator equation $\tilde L_1 {\cal D}= {\cal D}L_1$, where $L_1$ is defined in (\ref{SDYM}), $\tilde L_1=\partial_{t_1}+\lambda\partial_{z_1}+\tilde U_{z_1}$, and ${\cal D}$ is the unknown Darboux operator. 

These elementary considerations give a simple explanation of the difficulty in constructing Darboux - Backl\"und transformations of integrable dispersionless PDEs. 

If the Lax pair is made of vector fields containing the operator $\partial_{\lambda}$, like in the examples (\ref{dKP-system}), (\ref{dKP}), no results are known so far concerning the recursion operator generating the hierarchy of commuting flows. However, elegant alternative characterizations of such hierarchies in the case of Hamiltonian vector fields can be found, f.i., in \cite{Kri1}-\cite{TT3}, and, in the general case, f.i., in \cite{Bogdanov1,Bogdanov2}.

\end{document}